\documentclass[conference]{IEEEtran}
\usepackage{cite}
\usepackage{amsmath}
\usepackage{amssymb}
\usepackage{latexsym}
\usepackage{float}
\usepackage{epsfig}
\usepackage{subfigure,epstopdf,graphicx,array,algorithm}
\usepackage[]{graphicx}
\usepackage{algpseudocode}
\usepackage{hyperref}
\ifCLASSINFOpdf

\else
  
\fi

\begin{document}

\title{Improved Multiple Feedback Successive Interference Cancellation Algorithm for Near-Optimal MIMO Detection}

\author{\IEEEauthorblockN{Manish~Mandloi, Mohammed~Azahar~Hussain and Vimal~Bhatia}
\IEEEauthorblockA{Discipline of Electrical Engineering, Indian Institute of Technology Indore\\
Indore, Madhya Pradesh-453446\\
Email:\{phd1301202006, mt1402102007, vbhatia\}@iiti.ac.in}}

\maketitle

\begin{abstract} 
In this article, we propose an improved multiple feedback successive interference cancellation (IMF-SIC) algorithm for symbol vector detection in multiple-input multiple-output (MIMO) spatial multiplexing systems. The multiple feedback (MF) strategy in successive interference cancellation (SIC) is based on the concept of shadow area constraint (SAC) where, if the decision falls in the shadow region multiple neighboring constellation points will be used in the decision feedback loop followed by the conventional SIC. The best candidate symbol from multiple neighboring symbols is selected using the maximum likelihood (ML) criteria. However, while deciding the best symbol from multiple neighboring symbols, the SAC condition may occur in subsequent layers which results in inaccurate decision. In order to overcome this limitation, in the proposed algorithm, SAC criteria is checked recursively for each layer. This results in successful mitigation of error propagation thus significantly improving the bit error rate (BER) performance. Further, we also propose an ordered IMF-SIC (OIMF-SIC) where we use log likelihood ratio (LLR) based dynamic ordering of the detection sequence. In OIMF-SIC, we use the term dynamic ordering in the sense that the detection order is updated after every successful decision. Simulation results show that the proposed algorithms outperform the existing detectors such as conventional SIC and MF-SIC in terms of BER, and achieves a near ML performance.
\end{abstract}


\IEEEpeerreviewmaketitle

\section{Introduction}
\label{sec1}
With the rapid growth in demand for higher data rates, wireless technology is shifting towards the systems with multiple antennas. Multiple-input multiple-output (MIMO) systems in wireless communications provide significant improvements in wireless link reliability and the achievable capacity \cite{ref1,ref2,ref3}. Recently, the systems with large number of antennas (tens to hundreds of antennas) called as \textit{large-MIMO} systems are getting increased attention \cite{ref4} because they provide additional multiplexing and diversity gains. Using spatial multiplexing in MIMO systems, multiple data streams can be transmitted simultaneously from different transmit antennas. However, the detection of these data streams at the receiver end is challenging. To achieve minimum bit error rate (BER) performance in MIMO systems, an exhaustive search over all the possible transmit vectors is required. This is called as the maximum likelihood (ML) search for symbol vector detection in MIMO systems. However, as the number of antennas grow, the number of possible transmit vectors also increases exponentially and thus the ML search becomes computationally impractical. Sphere decoder, a well known MIMO detector achieves near ML performance but is practical only up to limited number of dimensions \cite{ref5}.\\
\indent The traditional MIMO detection techniques involve linear detection such as zero forcing (ZF) detector and minimum mean squared error (MMSE) detector. These detectors use linear transformation of the received vector in order to estimate the transmitted symbol vector. Comparatively the  MMSE detector is superior over ZF detector in terms of BER performance, but still their performance is far inferior compared to the ML performance. Another type of detection involve non-linear detectors such as vertical Bell labs layered space time architecture (V-BLAST) \cite{ref6} which utilizes ordered SIC for symbol vector detection. Some of the other MIMO detection technique involve ant colony optimization based MIMO detection \cite{ref7}, message passing based algorithm \cite{ref8}, channel hardening based algorithm \cite{ref9}, and lattice reduction aided MIMO detection \cite{ref10,ref11}. As an alternative, in this study we employ SIC based detection technique for MIMO detection problem.\\
\indent SIC based MIMO detection is a well known detection technique where symbols are detected sequentially \cite{ref12}. In SIC, after detecting each symbol, its interference is canceled from the received vector in order to improve the instantaneous signal to interference plus noise ratio (SINR) for the remaining symbols. It is also known as layered detection where in each layer, one symbol is detected. However, it suffers from error propagation which occur due to wrong decisions in early stages of the algorithm \cite{ref13, ref14}. For improving the performance of SIC, a well known technique known as V-BLAST \cite{ref6} technique is used which performs the following steps: (1) SNR based ordering in the detection sequence, (2) using ZF or MMSE for nulling the interference among the data streams and (3) detecting a symbol and canceling its interference i.e. SIC. Further, to mitigate the effect of error propagation, one of the available algorithms in the literature include multi-branch SIC (MB-SIC) algorithm \cite{ref15} where the concept of multiple branch (MB) processing is utilized. Each branch of MB-SIC differs in ordering of the detection sequence thus resulting in a higher detection diversity over the conventional SIC. Recently, a multiple feedback (MF) strategy is proposed in \cite{ref16} for SIC based MIMO detection. Multiple feedback SIC (MF-SIC) algorithm proposed in \cite{ref16} is based on the concept of shadow are constraint (SAC) where if the decision fall in the shadow region (unreliability region) then multiple constellation points are used in the decision feedback loop. The best symbol from the multiple symbols used in the decision feedback is selected using the ML criteria.\\
\indent In this article, we propose an improved multiple feedback successive interference cancellation (IMF-SIC) for symbol vector detection in MIMO spatial multiplexing systems. In MF-SIC, once a decision falls into the shadow region, conventional SIC is used to find the best one symbol for the corresponding layer. However, it may happen that while computing the best symbol for a given layer, multiple decisions from the subsequent layers fall into the shadow region. This condition is not checked in MF-SIC which sometimes results in error propagation and limits the BER performance. In improved MF-SIC (IMF-SIC), we overcome this limitation by checking the shadow region criteria recursively rather than using the conventional SIC for searching the best candidate symbol. This results in significant reduction in error propagation and thus the BER performance can be improved. Also, we propose an ordered IMF-SIC (OIMF-SIC) algorithm where we employ the log likelihood ratio (LLR) based dynamic ordering in the detection sequence \cite{ref17}. By the term dynamic ordering we mean that after every successful decision about a symbol, the ordering of the detection sequence is updated based on new LLR values. In OIMF-SIC, same ordering pattern is followed even for searching the best symbol when a decision is found unreliable. Simulation results show that the proposed algorithms significantly outperform the conventional SIC and the MF-SIC based detection methods, and that they achieve a near ML performance.\\
\indent Rest of the article is organized as follows. In Sect.~\ref{sec2}, we present mathematical model of MIMO system. An overview of some of the traditional MIMO detection methods is given in Sect.~\ref{sec3}. The proposed algorithms for symbol vector detection in MIMO spatial multiplexing systems is discussed in Sect.~\ref{sec4}. In Sect.~\ref{sec5}, we show the simulation results on comparison of BER performance. Finally, in Sect.~\ref{sec6} we conclude the article.

\section{System Model}
\label{sec2}
\begin{figure}
\centering
\includegraphics[width=8 cm ,height=5 cm]{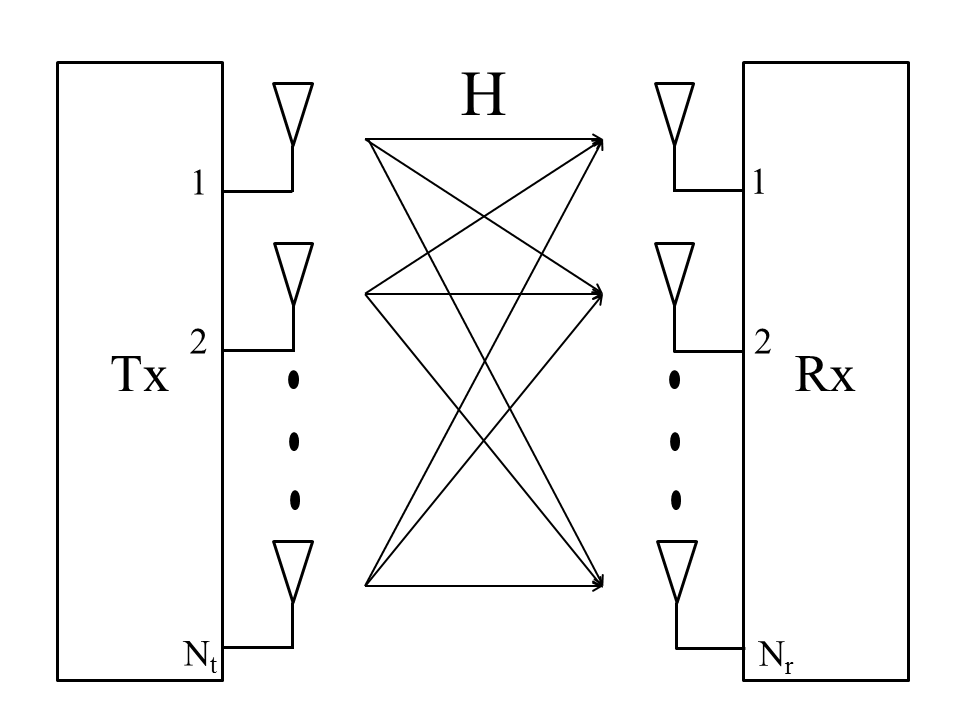}
\caption{MIMO system model.}
\label{figs1}
\end{figure}
In this section, we discuss the mathematical model of MIMO system as shown in Fig.\ref{figs1}. We consider a point to point spatially multiplexed MIMO link with $N_t$ transmit antennas and $N_r$ receive antennas. In spatial multiplexing, each antenna transmits a different information symbol $s_i$ for $i=1,2,\hdots,N_t$ . Let $\mathbf{s}=[s_1,  s_2,\cdots, s_{N_t}]^T$ be the $N_t\times1$ dimensional transmit vector, $[\cdot]^T$ denote the transpose of a matrix. Each entry $s_i$ of the transmit vector $\mathbf{s}$ is taken from a signal constellation $\mathbb{A}$, for example $\mathbb{A}=\{-1-1i,\ -1+1i,\ 1-1i,\ 1+1i\}$ for 4-QAM signaling. The channel over which the symbol vector $\mathbf{s}$ is transmitted is assumed to be a Rayleigh distributed flat fading channel. Let us consider an $N_r\times N_t$ complex channel matrix $\mathbf{H}$ with its elements $h_{i,j}$ for $i=1,2,\hdots,N_r$ and $j=1,2,\hdots,N_t$ assumed to be independent and identically distributed (i.i.d.) as complex normal with mean 0 and variance 1 i.e. $\sim\mathcal{CN}(0,1)$. The received symbol vector $\mathbf{y}$, after demodulation and matched filtering can be written as
\begin{equation}
\label{eq1}
\mathbf{y}=\mathbf{H}\mathbf{s}+\mathbf{n},
\end{equation}
where $\mathbf{n}$ is the additive white Gaussian noise (AWGN) with its element $n_i$ for $i=1,2,\hdots,N_r$ as i.i.d.  $\sim\mathcal{CN}(0,\sigma^2)$. The signal to noise ratio (SNR) is defined as $10\log_{10}\frac{N_tE_s}{\sigma^2}$, where $E_s$ is the average energy per symbol defined as $N_tE_s=\text{tr}(\mathbb{E}[\mathbf{s}\mathbf{s}^H])$ and $\sigma^2$ is the element-wise noise variance. tr($\cdot$) denote the trace of a matrix, $\mathbb{E}[\cdot]$ denote the expectation operation and $(\cdot)^H$ denote the Hermitian of a matrix. The channel state information (CSI) is assumed to be perfectly known at the receiver but is unknown at the transmitter. The ML solution for a given received vector $\mathbf{y}$ and a known channel matrix $\mathbf{H}$ can be found as
\begin{equation}
\label{eq2}
\mathbf{s}_{ML}=\arg\min_{\mathbf{s}\in\mathbb{A}^{N_t}}\Vert{\mathbf{y}-\mathbf{H}\mathbf{s}}\Vert^2.
\end{equation}
ML detection (MLD) requires an exhaustive search over $M^{N_t}$ possible transmit symbol vectors where $M$ is cardinality of the constellation set $\mathbb{A}$. This search grows exponentially with increase in the modulation order ($M$) or increase in the number of transmit antennas ($N_t$) or both.

\section{Overview of traditional MIMO detection techniques}
\label{sec3}
In this section, we discuss some of the traditional MIMO detection techniques such as linear detection techniques which involve zero forcing (ZF) based MIMO detection and minimum mean squared error (MMSE) based MIMO detection, and nonlinear detection techniques such as SIC based MIMO detection.
\subsection{Zero Forcing based MIMO detection}
\label{ssec31}
In ZF based MIMO detection, a linear transformation matrix $\mathbf{W}_{ZF}$ is used in such a way that it removes the effect due to the channel impairments on the received vector $\mathbf{y}$ i.e. $\mathbf{W}_{ZF}=(\mathbf{H}^H\mathbf{H})^{-1}\mathbf{H}^H$ which is the pseudo inverse of channel matrix $\mathbf{H}$. The linear transformation of the received vector using $\mathbf{W}_{ZF}$ can be written as
\begin{equation}
\label{eq3}
\widetilde{\mathbf{y}}=\mathbf{W}_{ZF}\mathbf{y}=\mathbf{s}+\mathbf{W}_{ZF}\mathbf{n},
\end{equation}
Thus the ZF estimate of the transmitted symbol can be found as
\begin{equation}
\label{eq4}
\mathbf{s}_{ZF}=\mathcal{Q}[\mathbf{W}_{ZF}\mathbf{y}],
\end{equation}
where $\mathcal{Q}[\cdot]$ is the quantization operation which maps the soft values on to the nearest constellation point. The computational complexity of ZF receiver is very less, however the BER performance is far inferior to the ML performance. The main drawback of ZF based MIMO detection is the problem of noise enhancement which degrades its performance.

\subsection{MMSE based MIMO detection}
\label{ssec32}
MMSE based MIMO detection is based on finding a suitable linear transformation matrix which minimizes the mean squared error between the transmit vector $\mathbf{x}$ and the transformed received vector as
\begin{equation}
\label{eq5}
\mathbf{W}_{MMSE}=\min_{\mathbf{W}}\mathbb{E}[\Vert{\mathbf{x}-\mathbf{W}\mathbf{y}}\Vert^2].
\end{equation}
The solution to Eq. \ref{eq5} can be written as
\begin{equation}
\label{eq6}
\mathbf{W}_{MMSE}=(\mathbf{H}^H\mathbf{H}+\frac{\sigma^2}{E_s}\mathbf{I}_{N_r})^{-1}\mathbf{H}^H.
\end{equation}
Using the transformation matrix $\mathbf{W}_{MMSE}$, the MMSE solution can then be computed as
\begin{equation}
\label{eq7}
\mathbf{s}_{MMSE}=\mathcal{Q}[\mathbf{W}_{MMSE}\mathbf{y}].
\end{equation}
MMSE based MIMO detection overcomes the problem of noise enhancement in ZF detectors. Thus it achieves better BER performance over ZF but still the performance of MMSE receiver is inferior as compared to the ML performance. One more drawback is that the MMSE solution needs knowledge of the noise variance $\sigma^2$.
\subsection{SIC based MIMO detection}
\label{ssec33}
In SIC based MIMO detection, the symbols transmitted from different transmit antennas are detected in a sequential manner. After detecting a symbol for a particular transmit antenna, its interference is canceled from the received vector in order to reduce the interference between different data streams at the receiver. The received vector in Eq. \ref{eq1} can be rewritten as
\begin{equation}
\label{eq8}
\mathbf{y}=\left(\mathbf{h}_1s_1+\mathbf{h}_2s_2+\cdots+\mathbf{h}_{N_t}s_{N_t}\right)+\mathbf{n}.
\end{equation}
Using the knowledge of channel matrix $\mathbf{H}$ a filter $\mathbf{w}_i$ is defined as
\begin{equation}
\label{eq9}
\mathbf{w}_i=(\mathbf{H}_{i-1}\mathbf{H}_{i-1}^H+\frac{\sigma^2}{E_s}\mathbf{I_{N_r}})^{-1}\mathbf{h}_i,
\end{equation}
where $\mathbf{H}_{i-1}$ is the matrix with its column taken from the $i$ to $N_t$ columns of the channel matrix $\mathbf{H}$. We use vector $\mathbf{y}_i$ as the vector left after removing the interference due to $s_1, s_2,\hdots, s_{i-1}$ as
\begin{equation}
\label{eq10}
\mathbf{y}_i=\mathbf{y}-\sum_{j=1}^{i-1}\mathbf{h}_js_j.
\end{equation}
Using $\mathbf{w}_i$, the first symbol $s_i$ is estimated as
\begin{equation}
\label{eq11}
\hat{s}_i=\mathcal{Q}[\widetilde{z}_i]=\mathcal{Q}[\mathbf{w}_i^H\mathbf{y}_i],
\end{equation}
where $\hat{s}_i$ is estimate of the transmitted symbol $s_i$. $\mathbf{w}_i$, $\mathbf{H}_i$ and $\mathbf{y}_i$ are updated after every successful decision about a symbol. SIC is an iterative detection technique, where in every iteration one symbol is detected. Thus a total of $N_t$ iterations are performed to detect the complete symbol vector. A pseudo code of SIC based MIMO detection is given in Algorithm \ref{algo1}.
\begin{algorithm}[H]
{
\caption{SIC based MIMO detection}
\label{algo1}
\begin{algorithmic}[1]
\State {{\bf input:} ${\bf y}$, ${\bf H}$, $N_t, N_{r}$};
\State {\bf initialize: $\mathbf{y_0=\mathbf{y}}, \mathbf{H}_0=\mathbf{H}$} ;
\For {$i=1:1:N_t$}
\State Compute $\mathbf{w}_i=(\mathbf{H}_{i-1}\mathbf{H}_{i-1}^H+\frac{\sigma^2}{E_s}\mathbf{I_{N_r}})^{-1}\mathbf{h}_i$
\State $\widetilde{z}_i=\mathbf{w}_i^H\mathbf{y}_{i-1}$
\State $\hat{s}_i=\mathcal{Q}[\widetilde{z}_i]$
\State Update $\mathbf{y}_i=\mathbf{y}_{i-1}-\mathbf{h}_i\hat{s}_i$
\State Update $\mathbf{H}_i=[\mathbf{h}_{i+1} \mathbf{h}_{i+2} \cdots \mathbf{h}_{N_t}]$
\EndFor
\State {{\bf output:} $\hat{\mathbf{s}}=[\hat{s}_1 \hat{s}_2 \cdots \hat{s}_{N_t}]$ is output solution vector}
\end{algorithmic}
}
\end{algorithm}
\indent However, SIC suffers from the problem of error propagation which occurs due to the wrong decisions in early stage of the algorithm. Once an error occurs, it propagates to the later stages and thus increases the number of errors. This degrades the BER performance of SIC based MIMO detection.
\section{Proposed algorithm for MIMO detection}
\label{sec4}
In this section, first we discuss the MF-SIC based algorithm for symbol vector detection in MIMO systems. We will then discuss the IMF-SIC algorithm and OIMF-SIC algorithm. 
\subsection{MF-SIC algorithm}
\label{ssec41}
The conventional SIC suffers from the problem of error propagation which degrades the BER performance. As a solution to mitigate the effect of error propagation, MF strategy has been proposed in \cite{ref16} for SIC based MIMO detection. The MF strategy is based on the concept of reliability of a soft decision in SIC. Let us consider the $k^{th}$ soft decision within SIC as
\begin{equation}
\label{eq12}
\tilde{z}_k=\mathbf{w}_k^H\mathbf{y}_k.
\end{equation}
The constellation point near to the soft decision $\tilde{z}_k$ is nothing but its quantized value i.e. $\mathcal{Q}[\tilde{z}_k]$. Let us define the distance between the soft decision and its quantized value as
\begin{equation}
\label{eq13}
d=\vert\tilde{z}_k-\mathcal{Q}[\tilde{z}_k]\vert
\end{equation}
If the soft decision in SIC is within a predefined threshold radius $d_{th}$ (i.e. if $d\leq d_{th}$) around the nearest constellation point then the decision is said to be reliable and its quantized value is used in the decision feedback. But if the decision is not within the radius (if $d>d_{th}$) then multiple neighboring (say $S$) constellation points are used instead of the quantized value followed by the conventional SIC. It can be thought of as running parallel streams one for each symbol and at last the best one is selected using the ML metric (Eq. \ref{eq2}). MF-SIC performs superior than the conventional SIC in terms of BER. A pseudo code of MF-SIC is shown in Algorithm \ref{algo2}
\begin{algorithm}[H]
{
\caption{MF-SIC based MIMO detection}
\label{algo2}
\begin{algorithmic}[1]
\State {{\bf input:} ${\bf y}$, ${\bf H}$, $N_t, N_{r}, d_{th}, S$};
\State {\bf initialize: $\mathbf{y_0=\mathbf{y}}, \mathbf{H}_0=\mathbf{H}$} ;
\For {$i=1:1:N_t$}
\State Compute $\mathbf{w}_i=(\mathbf{H}_{i-1}\mathbf{H}_{i-1}^H+\frac{\sigma^2}{E_s}\mathbf{I_{N_r}})^{-1}\mathbf{h}_i$
\State $\widetilde{z}_i=\mathbf{w}_i^H\mathbf{y}_{i-1}$
\State $d=\vert\widetilde{z}_i-\mathcal{Q}[\widetilde{z}_i]\vert$
\If {$d\leq d_{th}$}
\State $\hat{s}_i=\mathcal{Q}[\widetilde{z}_i]$
\Else
\For {$j=1:1:S$}
\State $x_i^{(j)}=\mathcal{N}^{(j)}(\tilde{z}_i)$ i.e. initialize by $j^{th}$ neighborhood of $\tilde{z}_i$
\For {$k=i+1:1:N_t$}
\State $\widetilde{z}_k=\mathbf{w}_k^H\mathbf{y}_{k-1}$
\State $\hat{s}_k=\mathcal{Q}[\widetilde{z}_k]$
\State Update $\mathbf{y}_k=\mathbf{y}_{k-1}-\mathbf{h}_k\hat{s}_k$
\State Update $\mathbf{H}_k=[\mathbf{h}_{k+1} \mathbf{h}_{k+2} \cdots\mathbf{h}_{N_t}]$
\EndFor
\State Compute the solution vector $\mathbf{x}^{(j)}$
\EndFor
\State $j_{opt}=\arg\min_{j=1,\hdots,S}{\Vert\mathbf{y}-\mathbf{H}\mathbf{x}^{(j)}\Vert^2}$
\State $\hat{s}_i=x_i^{(j_{opt})}$
\EndIf
\State Update $\mathbf{y}_i=\mathbf{y}_{i-1}-\mathbf{h}_i\hat{s}_i$
\State Update $\mathbf{H}_i=[\mathbf{h}_{i+1} \mathbf{h}_{i+2} \cdots \mathbf{h}_{N_t}]$
\EndFor
\State {{\bf output:} $\hat{\mathbf{s}}=[\hat{s}_1 \hat{s}_2 \cdots \hat{s}_{N_t}]$ is output solution vector}
\end{algorithmic}
}
\end{algorithm}
\subsection{Proposed IMF-SIC algorithm}
\label{ssec42}
In IMF-SIC, we propose an improved MF strategy for SIC based symbol vector detection in MIMO spatial multiplexing systems. In MF-SIC algorithm \cite{ref16}, multiple feedback loops are followed by the conventional SIC to check for the optimal decision for a particular layer. However, while searching for the optimal decision, the Shadow region (unreliability) criteria is not considered in the lower layers and hence there are chances of selecting a sub-optimal decision about a symbol which degrades the BER performance. In order to improve the accuracy about a decision among multiple symbols in the feedback loop, we check the shadow region criteria recursively in IMF-SIC algorithm. We therefore formulate a separate sub-routine (function) named as imf\_sic in Algorithm \ref{algo3} which checks the unreliability condition and optimally decide the symbol. Sub-routine imf\_sic is called from the main function which is given in Algorithm \ref{algo4} and also from the sub-routine itself. We denote the number of calls of the sub-routine by $L$ in IMF-SIC algorithm which is nothing but the number of times unreliability condition is checked once a decision falls into the shadow region. The number of neighboring symbol which are used in the decision feedback is denoted by $S$.\\
\begin{algorithm}[H]
{
\caption{IMF-SIC for MIMO detection}
\label{algo3}
\begin{algorithmic}[1]
\State {{\bf input:} ${\bf y}$, ${\bf H}$, $N_t, N_{r}, d_{th}, S$};
\State {\bf initialize: $\mathbf{y_0=\mathbf{y}}, \mathbf{H}_0=\mathbf{H}$} ;
\For {$i=1:1:N_t$}
\State Compute $\mathbf{w}_i=(\mathbf{H}_{i-1}\mathbf{H}_{i-1}^H+\frac{\sigma^2}{E_s}\mathbf{I_{N_r}})^{-1}\mathbf{h}_i$
\State $\widetilde{z}_i=\mathbf{w}_i^H\mathbf{y}_{i-1}$
\State $d=\vert\widetilde{z}_i-\mathcal{Q}[\widetilde{z}_i]\vert$
\If {$d\leq d_{th}$}
\State $\hat{s}_i=\mathcal{Q}[\widetilde{z}_i]$
\Else
\State $\hat{s}_i=$ imf\_sic($\mathbf{H}_{i-1},d_{th}, \mathbf{y}_{i-1}, i$)
\EndIf
\State Update $\mathbf{y}_i=\mathbf{y}_{i-1}-\mathbf{h}_i\hat{s}_i$
\State Update $\mathbf{H}_i=[\mathbf{h}_{i+1} \mathbf{h}_{i+2} \cdots \mathbf{h}_{N_t}]$
\EndFor
\State {{\bf output:} $\hat{\mathbf{s}}=[\hat{s}_1 \hat{s}_2 \cdots \hat{s}_{N_t}]$ is output solution vector}
\end{algorithmic}
}
\end{algorithm}

\begin{algorithm}[H]
{
\caption{imf\_sic}
\label{algo4}
\begin{algorithmic}[1]
\State function [$\hat{s}_i$]=imf\_sic($\mathbf{H}_{i-1},d_{th}, \mathbf{y}_{i-1}, i$)
\For {$j=1:1:S$}
\State $x_i^{(j)}=\mathcal{N}^{(j)}(\tilde{z}_i)$ i.e. initialize by $j^{th}$ neighborhood of $\tilde{z}_i$
\For {$k=i+1:1:N_t$}
\If {$d\leq d_{th}$}
\State $\hat{s}_i=\mathcal{Q}[\widetilde{z}_i]$
\Else
\State $\hat{s}_i=$ imf\_sic($\mathbf{H}_{k-1},d_{th}, \mathbf{y}_{k-1}, k$)
\EndIf
\State Update $\mathbf{y}_k=\mathbf{y}_{k-1}-\mathbf{h}_k\hat{s}_k$
\State Update $\mathbf{H}_k=[\mathbf{h}_{k+1} \mathbf{h}_{k+2} \cdots\mathbf{h}_{N_t}]$
\EndFor
\State Compute the solution vector $\mathbf{x}^{(j)}$
\EndFor
\State $j_{opt}=\arg\min_{j=1,\hdots,S}{\Vert\mathbf{y}-\mathbf{H}\mathbf{x}^{(j)}\Vert^2}$
\State $\hat{s}_i=x_i^{(j_{opt})}$
\State return $\hat{s}_i$
\end{algorithmic}
}
\end{algorithm}
\indent Significant reduction in the effect of error propagation is noticed using IMF-SIC (discussed in Sect. \ref{sec5}) which results in improvement in the BER performance over SIC and the MF-SIC. However, IMF-SIC does not include ordering in the detection sequence which plays a very important role in mitigating error propagation in SIC based detection techniques. In the next section, we discuss ordered IMF-SIC (OIMF-SIC) algorithm for MIMO detection.
\subsection{Proposed Ordered IMF-SIC (OIMF-SIC) algorithm}
\label{ssec43}
In this section, we propose an ordered IMF-SIC (OIMF-SIC) algorithm which incorporates the LLR based ordering in the detection sequence. Error propagation can be mitigated successfully if the decisions in the early layers of SIC are detected correctly. In several SIC based detection algorithms, channel norm ordering is used such as in V-BLAST technique \cite{ref6}. However, channel norm ordering does not consider the instantaneous noise in the received vector which has a significant impact on the BER performance. Thus a log likelihood ratio based ordering for multi-user detection in CDMA systems \cite{ref17} has been proposed in \cite{ref18} for MF-SIC based MIMO detection to overcome the sub-optimal behavior of channel norm based ordering. We employ LLR based dynamic ordering in the IMF-SIC algorithm to order the detection sequence. The term dynamic ordering is used in the sense that the ordering is updated after every successful symbol decision. However, the modification in the proposed method is that the ordering is also used in the oimf\_sic sub-routine which help in a more precise decision. The LLR value of a soft estimate $\tilde{z}_j$ can be written as \cite{ref18}
\begin{equation}
\label{eq14}
L_j=(1-\mathbf{h}_j^H\mathbf{R}^{-1}\mathbf{h}_j)^{-1}\vert\tilde{z}_j\vert
\end{equation}
where $\mathbf{R}=\mathbf{H}\mathbf{H}^H+\sigma^2\mathbf{I}$. The value of the argument $j$ which maximizes $L_j$ is arranged in descending order. Let us consider a set $B$ be the set of indices of the symbols which are not detected so far. The set $B$ is updated after every successful decision. For ordering the detection sequence in every iteration we consider the indices which are in the set $B$ i.e. in Eq. \ref{eq14} $j\in B$. Let us also consider the ordered sequence $\mathbf{t}$ which is updated in every iteration. In Algorithm \ref{algo5}, we present the OIMF-SIC algorithm for MIMO detection.
\begin{algorithm}[H]
{
\caption{OIMF-SIC for MIMO detection}
\label{algo5}
\begin{algorithmic}[1]
\State {{\bf input:} ${\bf y}$, ${\bf H}$, $N_t, N_{r}, d_{th}, S$};
\State {\bf initialize: $\mathbf{y_1=\mathbf{y}}, \mathbf{H}_1=\mathbf{H}, \mathbf{R}$ and $B = \{1, 2, \hdots, N_t\}$}
\For {$m=1:1:N_t$}
\For {$l=1:1:length(B)$}
\State Compute $\mathbf{w}_{b_l}=(\mathbf{H}_{b_l}\mathbf{H}_{b_l}^H+\frac{\sigma^2}{E_s}\mathbf{I_{N_r}})^{-1}\mathbf{h}_{b_l}$
\State $\widetilde{z}_{b_l}=\mathbf{w}_{b_l}^H\mathbf{y}_{b_l}$
\State Compute $L_{b_l}$ using the equation \ref{eq14}
\EndFor
\State Arrange the values $L_{b_j} \forall b_j \in B$ in descending order 
\State Update the detection order $\mathbf{t}$
\State Start detection using IMF-SIC given in Algorithm \ref{algo3}with ordered sequence $\mathbf{t}$.

\EndFor
\State {{\bf output:} $\hat{\mathbf{s}}=[\hat{s}_1 \hat{s}_2 \cdots \hat{s}_{N_t}]$ is output solution vector}
\end{algorithmic}
}
\end{algorithm}

\section{Simulation results}
\label{sec5}
\begin{figure}
\centering
\includegraphics[width=9 cm ,height=10 cm]{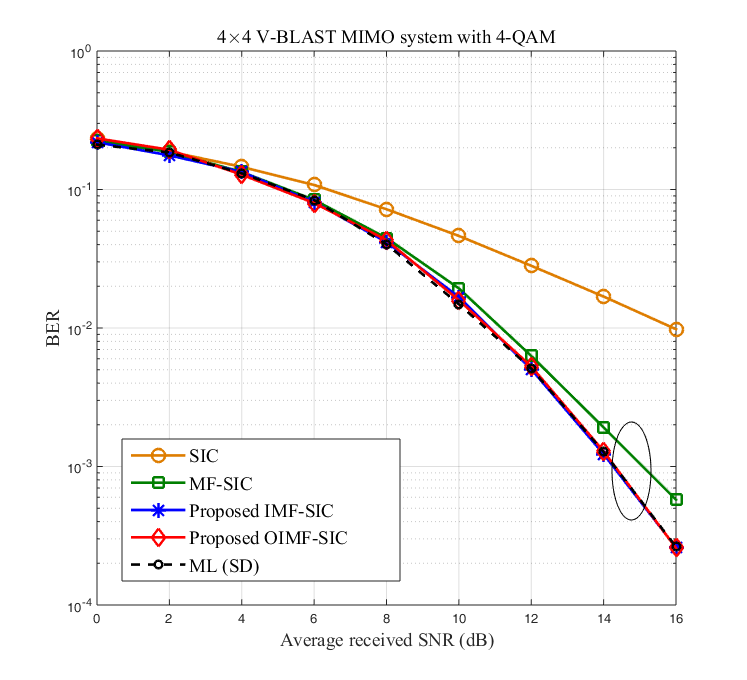}
\caption{BER performance comparison for uncoded $4\times 4$ V-BLAST MIMO system with 4-QAM and the value used for $d_{th}=0.2$, $L = 2$ and $S = 4$.}
\label{figs2}
\end{figure}

\begin{figure}
\centering
\includegraphics[width=9 cm ,height=10 cm]{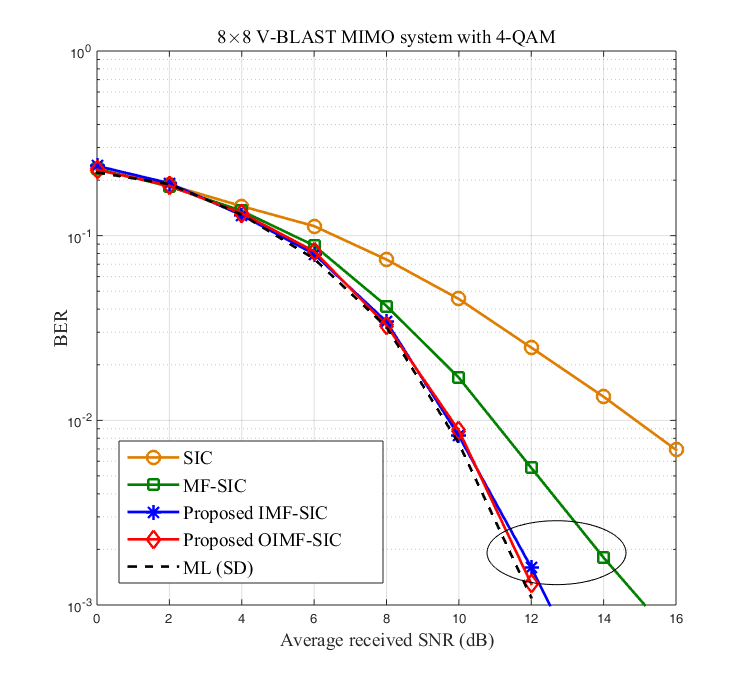}
\caption{BER performance comparison for uncoded $8\times 8$ V-BLAST MIMO system with 4-QAM and the value used for $d_{th}=0.2$, $L = 2$ and $S = 4$.}
\label{figs3}
\end{figure}

\begin{figure}
\centering
\includegraphics[width=9 cm ,height=10 cm]{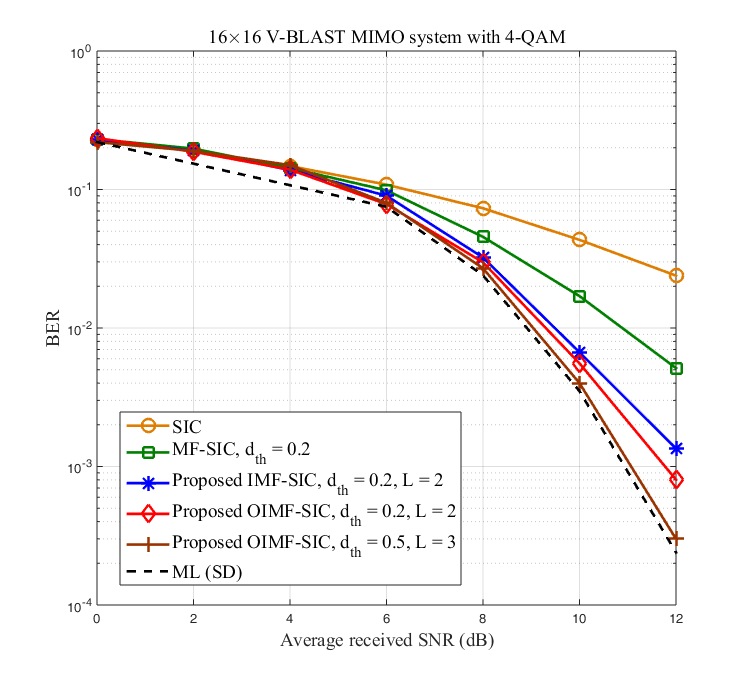}
\caption{BER performance comparison for uncoded $16\times 16$ V-BLAST MIMO system with 4-QAM.}
\label{figs4}
\end{figure}

\begin{figure}
\centering
\includegraphics[width=9 cm ,height=10 cm]{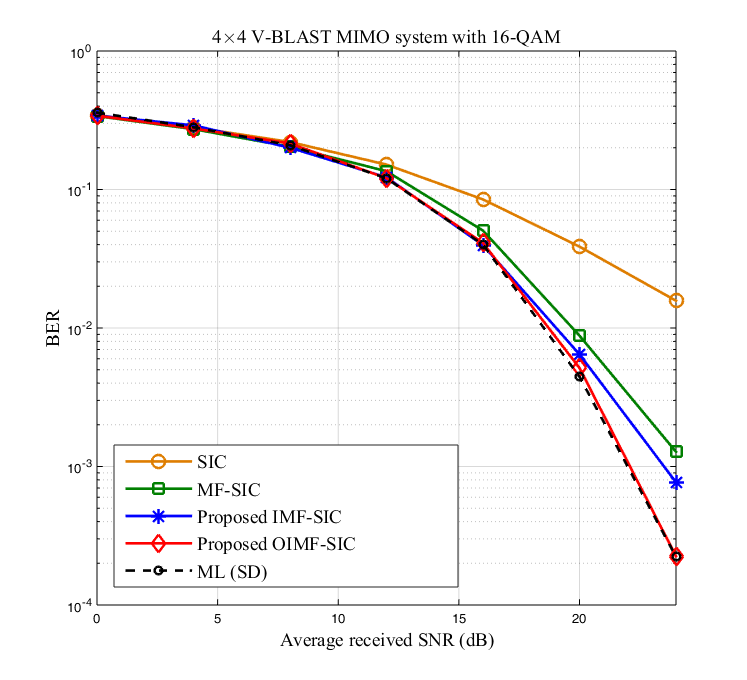}
\caption{BER performance comparison for uncoded $4\times 4$ V-BLAST MIMO system with 16-QAM and the value used for $d_{th}=0.2$, $L = 2$ and $S = 8$.}
\label{figs5}
\end{figure}

\begin{figure}
\centering
\includegraphics[width=9 cm ,height=10 cm]{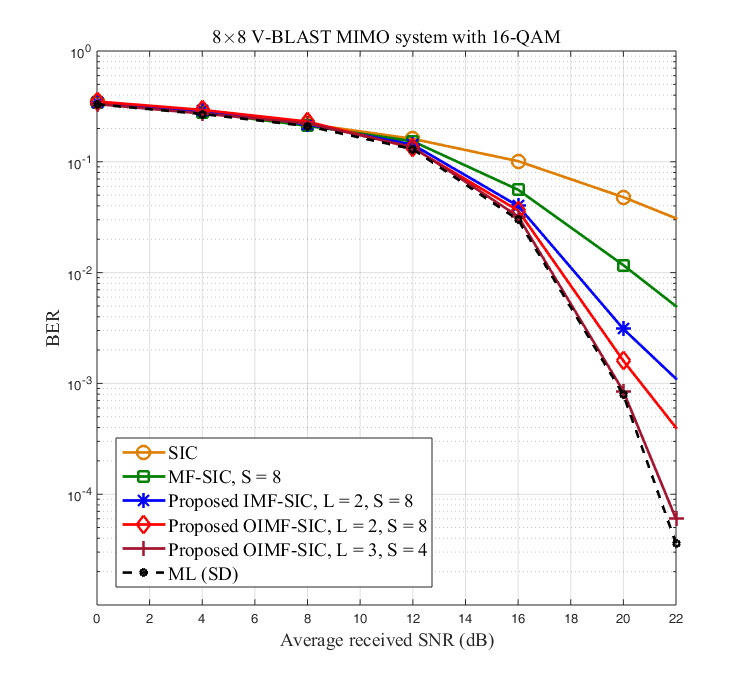}
\caption{BER performance comparison for uncoded $8\times 8$ V-BLAST MIMO system with 16-QAM.}
\label{figs6}
\end{figure}

In this section, we discuss the BER performance plots of the proposed algorithms and compare with the performance of the conventional SIC and the MF-SIC algorithm for $4\times 4$, $8\times 8$ and $16\times 16$ MIMO systems with 4-QAM and 16-QAM signaling. The results are generated using MATLAB and the BER is averaged over $10^4$ Monte-Carlo simulations.\\
\indent In Fig. \ref{figs2}, we compare the BER performance of the conventional SIC and MF-SIC with the proposed algorithms for $4\times 4$ MIMO system with 4-QAM. The value of threshold parameter $d_{th}$ used is 0.2 and the number of recursions $L=2$. Observation reveals that the proposed algorithms achieve a near ML performance and also the performance of IMF-SIC and OIMF-SIC algorithms is similar in terms of BER performance. At BER of $10^{-3}$, IMF-SIC results in approximately 1 dB gain in SNR over MF-SIC. In Fig. \ref{figs3}, the BER performance is compared for $8\times 8$ MIMO system with 4-QAM. A significant improvement in the BER performance of IMF-SIC and OIMF-SIC can be seen over SIC and MF-SIC based MIMO detection techniques. An SNR gain of around 2 dB can be achieved by using IMF-SIC over MF-SIC to achieve a target BER $2\times10^{-3}$. Due to near optimal performance of IMF-SIC for $4\times 4$ and $8\times 8$ MIMO systems with 4-QAM the advantage of dynamic ordering cannot be observed. However, with increase in number of antennas or the modulation order, ordering plays an important role in mitigating the error propagation.\\
\indent Fig. \ref{figs4} presents the BER performance comparison for $16\times 16$ V-BLAST MIMO system with 4-QAM. A significant improvement in the performance of OIMF-SIC can be seen when compared with IMF-SIC for $L=2$. However, to achieve the near ML performance the number of recursions are increased to $L=3$ and $d_{th}=0.5$ for OIMF-SIC. Thus, the effect of increase in number of recursions on the BER performance can be observed.\\
\indent The BER performance for $4\times 4$ and $8\times 8$ V-BLAST MIMO systems with 16-QAM signaling is shown in Figs. \ref{figs5} and \ref{figs6}. In 16-QAM the total number of constellation points is 16 and hence in order to get a reliable decision, we keep the number of neighboring contellation point in the feedback as $S=8$. The threshold value $d_{th}$ is kept same i.e. $d_{th}=0.2$. It can be observed that the BER performance of OIMF-SIC is close to within the 0.1 dB of the ML performance. Ordering in IMF-SIC results in almost 1 dB gain in SNR over the IMF-SIC without ordering. However, in Fig. \ref{figs6}, to achieve the BER close to ML, the number of recursions is increased to $L=3$ in OIMF-SIC. This shows a SNR gain of more than 2 dB over IMF-SIC at a target BER $10^{-3}$.

\section{Conclusions}
\label{sec6}
We proposed an improved multiple feedback successive interference cancellation (IMF-SIC) algorithm for symbol vector detection in MIMO systems. We also proposed an ordered IMF-SIC (OIMF-SIC) algorithm which employs log likelihood ratio (LLR) based dynamic ordering in the detection sequence. The ordering of the detection sequence is updated after every successful decision and thus the effect of error propagation is mitigated. The proposed algorithms significantly improves the BER performance and outperform the conventional SIC algorithm and the MF-SIC algorithm, and achieve near ML performance.
%
%

\section{*Acknowledgements}
The authors would like to thank IIT Indore and Ministry of Human Resource Development (MHRD) India for their support.


\end{document}